# Hybridization between the ligand *p* band and Fe-3*d* orbitals in the p-type ferromagnetic semiconductor (Ga,Fe)Sb

Takahito Takeda[1], Masahiro Suzuki[2], Le Duc Anh[1,3], Nguyen Thanh Tu[1,4], Thorsten Schmitt[5], Satoshi Yoshida[6], Masato Sakano[6], Kyoko Ishizaka[6], Yukiharu Takeda[7], Shin-ichi Fijimori[7], Munetoshi Seki[1,8], Hitoshi Tabata[1,8], Atsushi Fujimori[2,9], Vladimir N. Strocov[5], Masaaki Tanaka[1,8], and Masaki Kobayashi[1,8]

[1]*Department of Electrical Engineering and Information Systems, University of Tokyo, 7-3-1 Hongo, Bunkyo-ku, Tokyo 113-8656, Japan*

[2]*Department of Physics, University of Tokyo, 7-3-1 Hongo, Bunkyo-ku, Tokyo 113-0033, Japan*

[3]*Institute of Engineering Innovation. University of Tokyo, 7-3-1 Hongo, Bunkyo-ku, Tokyo 113-0033, Japan*

[4]*Department of Physics, Ho Chi Minh City University of Pedagogy, 280, Au Duong Vuong Street, District 5, Ho Chi Minh City 748242, Vietnam*

[5]*Swiss Light Source, Paul Scherrer Institut, CH-5232 Villigen PSI, Switzerland*

[6]*Quantum-Phase Electronics Center and Department of Applied Physics, University of Tokyo, Bunkyo, Tokyo 113-8656, Japan*

[7]*Synchrotron Radiation Research Unit, Japan Atomic Energy Agency, Sayo-gun, Hyogo 679-5148, Japan*

[8]*Center for Spintronics Research Network, University of Tokyo, 7-3-1 Hongo, Bunkyo-ku, Tokyo 113-8656, Japan*

[9]*Department of Applied Physics, Waseda University, Okubo, Shinjuku, Tokyo 169-8555, Japan*

(Date: January 9, 2019)

## ABSTRACT

(Ga,Fe)Sb is a promising ferromagnetic semiconductor for practical spintronic device applications because its Curie temperature ($T_C$) is above room temperature. However, the origin of ferromagnetism with high $T_C$ remains to be elucidated. Here, we use soft x-ray angle-resolved photoemission spectroscopy (SX-ARPES) to investigate the valence-band (VB) structure of $(Ga_{0.95},Fe_{0.05})Sb$ including the Fe-3*d* impurity band (IB), to unveil the mechanism of ferromagnetism in (Ga,Fe)Sb. We find that the VB dispersion in $(Ga_{0.95},Fe_{0.05})Sb$ observed by SX-ARPES is similar to that of GaSb, indicating that the doped Fe atoms hardly affect the band dispersion. The Fe-3*d* resonant ARPES spectra demonstrate that the Fe-3$d$ IB crosses the Fermi level ($E_\text{F}$) and hybridizes with the VB of GaSb. These observations indicate that the VB structure of (Ga$_{0.95}$,Fe$_{0.05}$)Sb is consistent with that of the IB model which is based on double-exchange interaction between the localized 3$d$ electrons of the magnetic impurities. The results indicate that the ferromagnetism in (Ga,Fe)Sb is formed by the hybridization of the Fe-3$d$ IB with the ligand $p$ band of GaSb.

## I. INTRODUCTION

Ferromagnetic semiconductors (FMSs) are alloy semiconductors in which cation sites are partially replaced by a sizable amount of magnetic impurities, leading to ferromagnetic properties. The ferromagnetism of FMSs is considered to originate from the magnetic interaction between the doped magnetic impurities mediated by the spin of the carriers. This nature is called carrier-induced ferromagnetism[1]. FMSs have attracted much attention as promising materials to apply for semiconductor spintronic devices, because one can control their magnetic properties by changing the carrier concentration. The p-type Mn-doped III-V FMSs, such as (In,Mn)As[2,3,4,5] and (Ga,Mn)As[6,7,8], have been intensively studied so far as prototypical FMSs showing carrier-induced ferromagnetism.

Spintronics devices based on these FMSs have been fabricated[9,10]. However, these materials are still seriously problematic for practical applications. Firstly, Mn-doped FMSs are only p-type because the doped Mn atoms act as acceptors in III-V semiconductors. Secondly, the reported maximum values of the Curie temperature ($T_C$) of molecular-beam epitaxy (MBE)-grown Mn-doped FMSs are much lower than room temperature. The highest $T_C$ of (Ga,Mn)As reported so far is ~ 200 K[11] and that of (In,Mn)As is ~ 90 K[12]. Recently, Fe-doped III-V FMSs such as n-type (In,Fe)As[13], n-type (In,Fe)Sb[14], and p-type (Ga,Fe)Sb[15] have been successfully grown. Since the doped Fe ions are expected to isovalently substitute for the cation ($In^{3+}$ or $Ga^{3+}$) sites of III-V semiconductors as $Fe^{3+}$, one can independently control the concentrations of Fe ions and, by doping other atoms, carriers in Fe-doped FMSs. Furthermore, the previously reported highest $T_C$ values n-type $(In_{0.84},Fe_{0.16})Sb$ (335 K)[16], $(In_{0.65},Fe_{0.35})Sb$ (385 K)[17], and p-type $(Ga_{0.8},Fe_{0.2})Sb$ (> 400 K)[18] are well above room temperature. Considering these advantages, Fe-based FMSs are more promising materials for applications to semiconductor spintronic devices operating at room temperature.

Understanding the mechanism of the carrier-induced ferromagnetism in FMSs is important for designing functional FMSs materials and for practical applications. Several theoretical models, such as the Zener's *p-d* exchange model and impurity band (IB) model have been proposed so far[19]. The former, based on a mean-field theory, has been proposed as the itinerant limit where the holes are considered to be nearly free carriers[20,21]. This model indicates that the Fermi level ($E_F$) is located in the valence band (VB) and that the origin of the ferromagnetism is the *p-d* exchange interactions between the VB holes and localized 3*d* electrons of the magnetic impurity. In contrast, the latter model has been proposed as the other limit where the hole carriers are localized around the magnetic impurities[22,23]. This model indicates that $E_F$ is located in the IB and that the ferromagnetism arises from the double-exchange interaction between the localized 3*d* electrons of the magnetic impurities. As described above, the origin of ferromagnetism in FMSs is considered to be related to its band structure near $E_F$.

Experimentally, the electronic states of (Ga,Mn)As in the vicinity of $E_F$ have been studied by angle-resolved photoemission spectroscopy (ARPES) to unveil the mechanism of carrier-induced ferromagnetism from the viewpoint of electronic band structure[24,25,26,27]. Soft x-ray (SX) ARPES measurements have been instrumental to directly access the three-dimensional (3D) band structure and Mn-3$d$ IB of (Ga,Mn)As[27].

Here, we investigate the electronic structure of the Fe-based FMS (Ga,Fe)Sb to reveal the origin of its high $T_C$, and particularly examine the Fe-3$d$ IB in the vicinity of $E_F$, by SX-ARPES measurements. The SX-ARPES results provide an understanding of the carrier-induced ferromagnetism of (Ga,Fe)Sb.

## II. EXPERIMENTAL

(Ga$_{0.95}$,Fe$_{0.05}$)Sb and GaSb thin films a thickness of 30 nm were grown on semi-insulating (SI) GaAs(001) substrates by MBE. To avoid surface contamination, the surfaces of the films were covered with amorphous Sb capping layers. Figure 1(a) shows the schematic sample structure. During the MBE growth, the excellent crystallinity of the samples was confirmed by reflection high-energy electron diffraction. The lattice constant ($a$) of (Ga$_{0.95}$,Fe$_{0.05}$)Sb is 0.608 nm [28]. The value of $T_C$ of (Ga$_{0.95}$,Fe$_{0.05}$)Sb, which was grown by the same condition as the measured one, is about 25 K[15]. The SX-ARPES experiments were performed at the SX-ARPES end station[29] of the ADRESS beamline at the Swiss Light Source. Before the SX-ARPES measurements, the samples were annealed at around 300 °C in the preparation chamber to remove the amorphous Sb capping layer and to expose the clean surface of the samples. The measurements were conducted under an ultrahigh vacuum below $10^{-10}$ mbar at a temperature of 12 K, with varying the photon energy ($hv$) from 500 eV to 1000 eV. The total energy resolution including the thermal broadening was between 50 meV and 200 meV depending on $hv$. The incident beam with linear-vertical and linear-horizontal polarizations, which respectively correspond to $p$-polarization and $s$-polarization configurations[30], were used for the measurements. The Fe $L_{2,3}$ x-ray absorption spectroscopy (XAS) spectra were measured in the total-electron-yield mode.

## III. RESULTS AND DISCUSSION

**A. Constant energy mappings and band dispersion around the Γ point**

Figures 2(a) - 2(c) show out-of-plane ($k_z$-$k_{//}$) constant-energy mappings for $(Ga_{0.95},Fe_{0.05})Sb$ at different binding energy ($E_B$), where the red solid lines represent the Brillouin zone (BZ) as shown Fig. 1(b). Here, $k_z$ and $k_{//}$ (= $k_{[-110]}$) are out-of-plane (Γ-X-Γ) and in-plane (Γ-K-X) momenta, respectively. The observed band dispersion clearly

depends on $k_z$ and reflects the symmetry of the BZ. Note here that there is no band that is non-dispersive along the $k_z$ direction in the mappings, evidencing the absence of surface (two-dimensional) states in the present SX-ARPES data. These observations demonstrate that the SX-ARPES spectra reflect the 3D band dispersion of the $(Ga_{0.95},Fe_{0.05})Sb$ thin film.

Figures 3(a) and 3(b) show the out-of-plane and in-plane Fermi surface mappings (FSMs) of the $(Ga_{0.95},Fe_{0.05})Sb$ thin film, respectively. It should be noted here that the cut taken at $hv = 885$ eV contains the Γ point, as shown in Fig. 3(a). Then, the Γ-K-X symmetry line is precisely determined from the in-plane $k_{[-110]}$-$k_{[-1-10]}$ constant-energy mapping taken at $hv = 885$ eV as shown in Fig. 3(b).

Figures 3(c) and 3(d) show $E_B$ vs. momentum ($k$) plots along the Γ-K-X symmetry line with $p$ and $s$ polarizations, respectively. The light-hole (LH) and split-off (SO) bands show up with $p$ polarization in Fig. 3(c), while the heavy-hole (HH) band is clearly visible in the SX-ARPES image taken with $s$ polarization [Fig. 3(d)]. This linear polarization dependence comes from the wave-function symmetry of these bands[31]. Following the LH-band peaks of energy distribution curves (EDCs) in Fig. 3(e), the top of the LH band is located at 150 meV below $E_F$. Because the band gap ($E_g$) of (Ga$_{0.95}$,Fe$_{0.05}$)Sb is close to that of GaSb ($E_g$ = 812 meV[32])[33], the observation evidences that (Ga$_{0.95}$,Fe$_{0.05}$)Sb is p-type, which is in agreement with the transport measurements[15,28,34].

To examine the Fe-doping effects on the band dispersion, ARPES measurements on the GaSb thin film has also been conducted as a reference. Figure 4(a) shows $E_B$ vs $k$ plots for GaSb along the Γ-K-X symmetry line with $p$ polarizations at the $hv$ of 880 eV. GaSb is p-type because the top of the LH band is close to $E_F$. Comparing the band dispersions between (Ga$_{0.95}$,Fe$_{0.05}$)Sb in Fig. 3(c) and GaSb in Fig. 4(a), the band dispersion of (Ga$_{0.95}$,Fe$_{0.05}$)Sb is nearly identical to that of GaSb. Figure 4(b) shows the EDCs of GaSb (blue line) and (Ga$_{0.95}$,Fe$_{0.05}$)Sb (red line) along the Γ-K-X symmetry line with $p$ polarization. The ARPES spectra of (Ga$_{0.95}$,Fe$_{0.05}$)Sb are broader than those of GaSb because of the structural disorder due to the Fe doping and/or because of the Zeeman splitting. However, the peak positions of EDCs of (Ga$_{0.95}$,Fe$_{0.05}$)Sb are almost the same as those of GaSb. This result indicates that the band dispersion itself originating from the *sp*-orbitals of GaSb is hardly affected by doping of Fe atoms. Since the position of $E_F$ in (Ga$_{0.95}$,Fe$_{0.05}$)Sb is approximately the same as that in GaSb, the Fe atoms would isovalently substitute for the Ga site.

**B. Fe-3$d$ impurity band**

Note that the Fe-3$d$ IB is hardly seen in the ARPES spectra of (Ga$_{0.95}$,Fe$_{0.05}$)Sb near the Γ point in Figs. 3(c) and 3(d) due to the small amount of Fe atoms, although the band dispersion of the GaSb host has been clearly observed. The energy position of the Fe-3$d$ states is a key to understand the ferromagnetism of (Ga$_{0.95}$,Fe$_{0.05}$)Sb because its position in the IB model is different from that in the $p$-$d$ Zener model. To determine the position of the Fe-3$d$ IB in the VB experimentally, we have conducted resonant angle-resolved photoemission spectroscopy (r-ARPES) at the Fe $L_3$ absorption edge. By using r-ARPES, the states derived from the orbitals which are relevant to the absorption process are resonantly enhanced, in our case the Fe $d$-states. Figure 5(a) shows the XAS spectrum at the Fe $L_3$ edge of the present (Ga$_{0.95}$,Fe$_{0.05}$)Sb film decapped by annealing at around 300 °C. Its shape is similar to that of the reported XAS spectra of the (Ga,Fe)Sb thin films capped by an amorphous As layer[35]. Since the XAS spectra of the capped (Ga,Fe)Sb thin films have been measured without annealing, this result indicates that the annealing hardly changes the 3$d$ states of Fe in the $(Ga_{0.95},Fe_{0.05})Sb$ film. A previous x-ray magnetic circular dichroism study of the capped (Ga,Fe)Sb thin films has demonstrated that the XAS peak of 708 eV originates from the ferromagnetic component[35]. The resonance enhancement of ARPES measured at this $hv$ should therefore reflect the energy position of the ferromagnetic component of Fe-3$d$ states in (Ga,Fe)Sb. Figures 5(b) and 5(c) show the r-ARPES images of the $(Ga_{0.95},Fe_{0.05})Sb$ thin films taken at $hv$ = 708 eV (on-resonance) and 704 eV (off-resonance) with $p$ polarizations. A flat band appears in the on-resonance spectrum in the vicinity of $E_F$ in Fig. 5(b), while this band apparently disappears in the off-resonance spectrum in Fig. 5(c). The Fe-3$d$ state in the vicinity of $E_F$ is consistent with the previous resonant photoemission spectroscopy (RPES) at the Fe $L_{2,3}$ edge of (Ga,Fe)Sb[35]. The flat band observed with $hv$ = 708 eV is therefore the Fe-3$d$ IB, which is related to the ferromagnetism.

Figure 5(d) shows the EDCs of the on- and off-resonance spectra from $k_{//}$ = 1.5 ($k_1$) to $k_{//}$ = 2.1 ($k_2$) in the reciprocal lattice unit of $2\sqrt{2}\pi/a$. The difference of the on-resonance EDCs from the off-resonance ones [the blue areas between the pink and green dashed curves in Fig. 5(d)] reveals the Fe-3$d$ component of the ARPES spectra. Here, it should be noted here that the on-resonant signal of the IB is much enhanced compared with its real contribution to the density of states (DOS). Generally, the states induced by impurities doped in a single crystal are independent of the wave number owing to the random atomic distribution. In our case, however, the Fe-3$d$ component areas strongly change with $k_\parallel$ along the VB dispersion as shown by the red dots in Figs. 5(b) and 5(d). This indicates that the intensity of this LH band is also enhanced at the Fe 2$p$-3$d$ resonance through hybridization between the Fe-3$d$ orbital and the ligand $p$ band[27]. Therefore, the present observation provides experimental evidence that the Fe-3$d$ IB hybridizes with the ligand $p$ band.

### C. Discussion

Finally, based on the experimental findings, we will discuss the band structure and the origin of the ferromagnetism of (Ga,Fe)Sb. From the observations described above, we have found that the Fe-3$d$ IB located just above the VB maximum (VBM) crosses $E_\mathrm{F}$ and hybridizes with the $p$ band of GaSb, and that the LH, HH and SO bands of ($Ga_{0.95}$,$Fe_{0.05}$)Sb similar to those of the host GaSb are located below $E_F$, as shown in Fig. 6(a). Figure 6(b) shows a schematic diagram of the electronic structure of ($Ga_{0.95}$,$Fe_{0.05}$)Sb.

The observation that the Fe-3$d$ IB shows the Fermi cutoff indicates that the Fe-3$d$ IB is partially occupied by electrons. Considering the result that the VBM is located below $E_F$, it is probable that the concentration of hole carriers due to defects, if they exist, is much lower than that of the Fe ions. Additionally, if the Fe ions are ionic $Fe^{3+}$, the majority(up)-spin states are fully occupied due to the $d^5$ electronic configuration. It follows from these arguments that the number of $d$ electrons in ($Ga_{0.95}$,$Fe_{0.05}$)Sb is expected to deviate from the half-filled $d^5$ configuration through the $p$-$d$ hybridization. The strength of the $p$-$d$ hybridization is closely related with the symmetry of $d$ electrons in (Ga,Fe)Sb. The five-fold degenerate state of Fe 3$d$ splits into the two-fold degenerate states ($e$) and the three-fold degenerate states ($t_2$) because the Fe ions of (Ga,Fe)Sb are in the tetrahedral crystal field. Due to the symmetries of the $e$ and $t_2$ states, while the $e$ states do not hybridize with the ligand $p$ bands, the $t_2$ states hybridize with the $p$ bands[36]. The $p$-$d(t_2)$ hybridization leads to the antibonding ($t_{2a}$) and bonding ($t_{2b}$) states. The $p$-$d(t_2)$ hybridization leads to the antibonding ($t_{2a}$) and bonding ($t_{2b}$) states, which have both the Fe $t_2$ and the ligand Sb $p$ characters. Figure 6 shows the schematic energy diagram of (Ga,Fe)Sb based on the experimental findings. Here, ↑ and ↓ means majority(up) spin and minority(down) spin, respectively. For the majority-spin states, since the Fe-$3d_\uparrow$ levels are located well below the VBM, the bonding $t_{2b\uparrow}$ and antibonding $t_{2a\uparrow}$ states have predominantly both the Fe $t_2$ and Sb $p$ characters, respectively. In contrast, since the Fe-$3d_\downarrow$ levels are located above the VBM, the minority-spin bonding $t_{2b\downarrow}$ state is mainly composed of the ligand $p$ orbitals, and the antibonding $t_{2a\downarrow}$ state primarily consists of the Fe $t_2$ orbitals. Band widths of the bonding $t_{2b}$ and antibonding $t_{2a}$ states with the $p$ characters will be comparable with or narrower than that of VB through the strong $p$-$d(t_2)$ hybridization due to the high covalency of the narrow-gap semiconductor GaSb. Considering that the VBM is located just below $E_F$, it is probable that the majority-spin antibonding $t_{2a\uparrow}$ state crosses $E_F$ and is partially filled (see the left side of Fig. 6(c)) because the $t_{2a\uparrow}$ is located above the VBM. To keep the charge balance of $Fe^{3+}$, the holes of the majority-spin $t_{2a\uparrow}$ state will be compensated by electron occupation of the minority-spin $e_\downarrow$ and/or $t_{2a\downarrow}$ states (see the right side of Fig. 6(c)). This is consistent with the first-principle calculations for (Ga,Fe)Sb that the majority-spin $t_{2a\uparrow}$ and the minority-spin $e_\downarrow$ and/or $t_{2a\downarrow}$ states cross $E_F$[37,38]. It follows from these arguments that the observed narrow Fe-3$d$ IB crossing $E_F$ probably originates from the partially occupied minority-spin $e_\downarrow$ and/or $t_{2a\downarrow}$ states, and the observed $p$-$d$ hybridized states in the vicinity of $E_F$ would be the $t_{2a\uparrow}$ and $t_{2b\downarrow}$ states with the primarily $p$ characters.

The present SX-ARPES study indicates that the electronic structure of (Ga$_{0.95}$,Fe$_{0.05}$)Sb is consistent with the IB model, where the Fe-3$d$ IB crosses $E_F$. This means that the carriers of (Ga,Fe)Sb are mainly derived from the 3$d$ electrons even if the $p$-$d$ hybridization is finite. It should be mentioned here that the transport properties of (Ga,Fe)Sb stay semiconducting when the Fe concentration is less than 20%[28], although the observed Fe-3$d$ IB of the (Ga$_{0.95}$,Fe$_{0.05}$)Sb thin film shows the Fermi cutoff. It is possible that either gap opening, or depletion of DOS near $E_F$ which is smaller than the experimental energy resolution occurs in the Fe-3$d$ IB of (Ga,Fe)Sb. Additionally, the actual contribution of the Fe-3$d$ IB to the transport properties is expected to be negligible because the Fermi cutoff of the Fe-3$d$ IB (or the spectral weight at $E_F$) is fairly small in the off-resonance spectra as shown in Fig. 5(b). The DOS of the Fe-3$d$ IB near $E_F$ will increase with increasing the Fe concentration, resulting in the increase of the conductivity and $T_C$ of (Ga,Fe)Sb with higher Fe concentrations. It follows from these arguments that the double-exchange interaction in the Fe-3$d$ IB crossing $E_F$ is the origin of the ferromagnetism in (Ga,Fe)Sb.

It has been reported that the local Fe concentration in (Ga,Fe)Sb fluctuates on the several-nm scale, and zinc-blende Fe-rich clusters are formed in the (Ga,Fe)Sb matrix [10, 22]. Since our (Ga,Fe)Sb thin film with $x$ = 0.05 is expected to have a less fluctuation of the local Fe concentration, further studies including a systemic SX-ARPES measurements of the VB structure of (Ga,Fe)Sb with varying $x$ is necessary to clarify the origin of the magnetic interaction. In addition, not only the conventional valence of Fe but also the intermediate valence of Fe should be taken into account because of the $p$-$d$ hybridization. To identify the local electronic structure of Fe ions including the electronic structure parameters such as the charge-transfer energy and strength of Coulomb interaction, it is indispensable to conduct other measurements sensitive to the local electronic structure, e.g., resonant inelastic x-ray scattering (RIXS) combined with cluster-model calculation. The effectiveness of RIXS for FMS has been confirmed in (Ga,Mn)As[39].

## IV. SUMMARY

We have performed SX-ARPES measurements on a $(Ga_{0.95},Fe_{0.05})Sb$ thin film to obtain the information about its VB structure, the location of the Fe-3$d$ IB, and its hybridization with the $p$ band. By capping samples with an amorphous Sb layer and removing the cap by annealing *in vacuo* just before performing measurements, we have succeeded in observing the 3D bulk VB structure in the samples with clean surfaces. Experimentally, the band dispersion of $(Ga_{0.95},Fe_{0.05})Sb$ is similar to that of GaSb. This indicates that the Fe ions hardly affect the band dispersion of the host GaSb. In addition, the non-dispersive Fe-3$d$ IB hybridized with the ligand $p$ bands is located around $E_F$, indicating that the carriers of (Ga,Fe)Sb have the $d$-like character. Based on our results, the electronic structure of (Ga,Fe)Sb is consistent with the IB model. It is probable that the partially filled Fe-3$d$ IB is composed of the minority-spin $e_\downarrow$ and/or $t_{2a\downarrow}$ states through the $p$-$d$ hybridization. Thus, double-exchange interaction between Fe ions would be the origin of the ferromagnetic interaction of (Ga,Fe)Sb. To thoroughly study the ferromagnetic mechanism of (Ga,Fe)Sb with higher Fe concentrations, systematic studies of the VB structure and the local electronic state of Fe-3$d$ electrons on the (Ga,Fe)Sb thin films with various Fe concentrations are needed.

## ACKNOWLEWDGMENTS

This work was supported by a Grant-in-Aid for Scientific Research (Grant Nos. 15H02109, 16H02095, 17H04922, 18H05345, and 23000010), Core-to-Core Program A. Advanced Research Networks from JSPS, and CREST of JST (No. JPMJCR1777), Japan. This work was partially supported the Spintronics Research Network of Japan (Spin-RNJ). Supporting experiments at SPring-8 were approved by the Japan Synchrotron Radiation Research Institute (JASRI) Proposal Review Committee (Proposal No. 2018A3841 and 2019A3841).

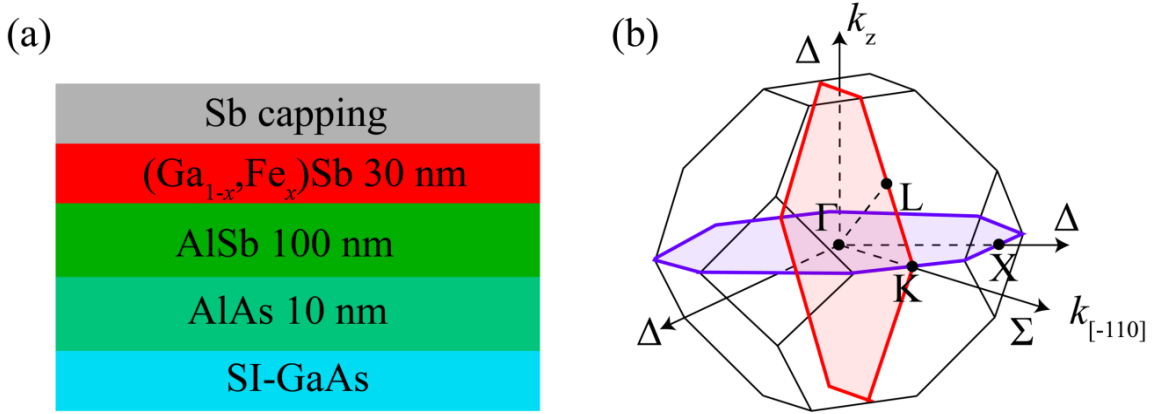

FIG. 1. Sample structure of $(Ga_{1-x},Fe_x)Sb$ thin films: (a) structure of the $(Ga_{1-x},Fe_x)Sb$ ($x$ = 0 and 0.05) thin films, (b) Brillouin zone (BZ) with the areas surrounded by red and purple lines measured in the out-of-plane and in-plane measurements, respectively.

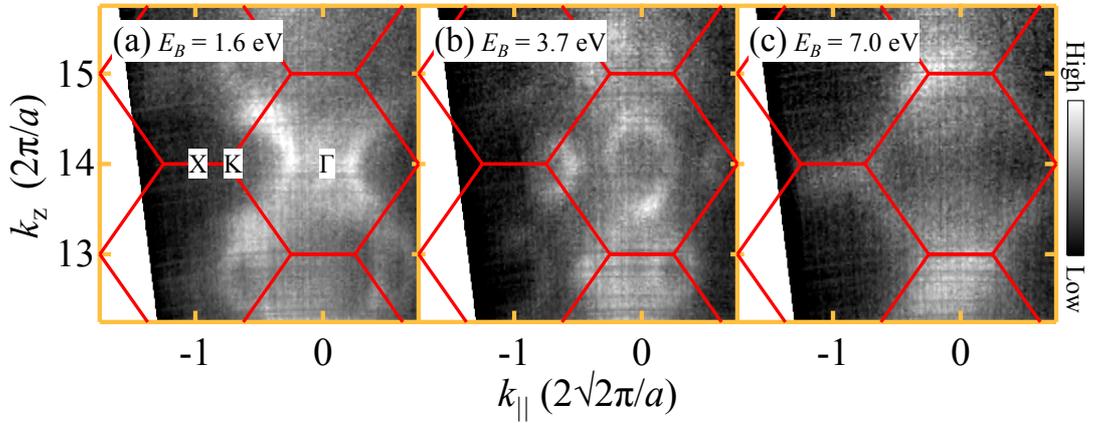

FIG. 2. Constant-energy mappings in the $k_z$-$k_\parallel$ plane: (a)-(c) mappings for $E_B$ = 1.6 eV, 3.7 eV, and 7.0 eV, respectively and red solid lines represent the BZ in Fig. 1(b).

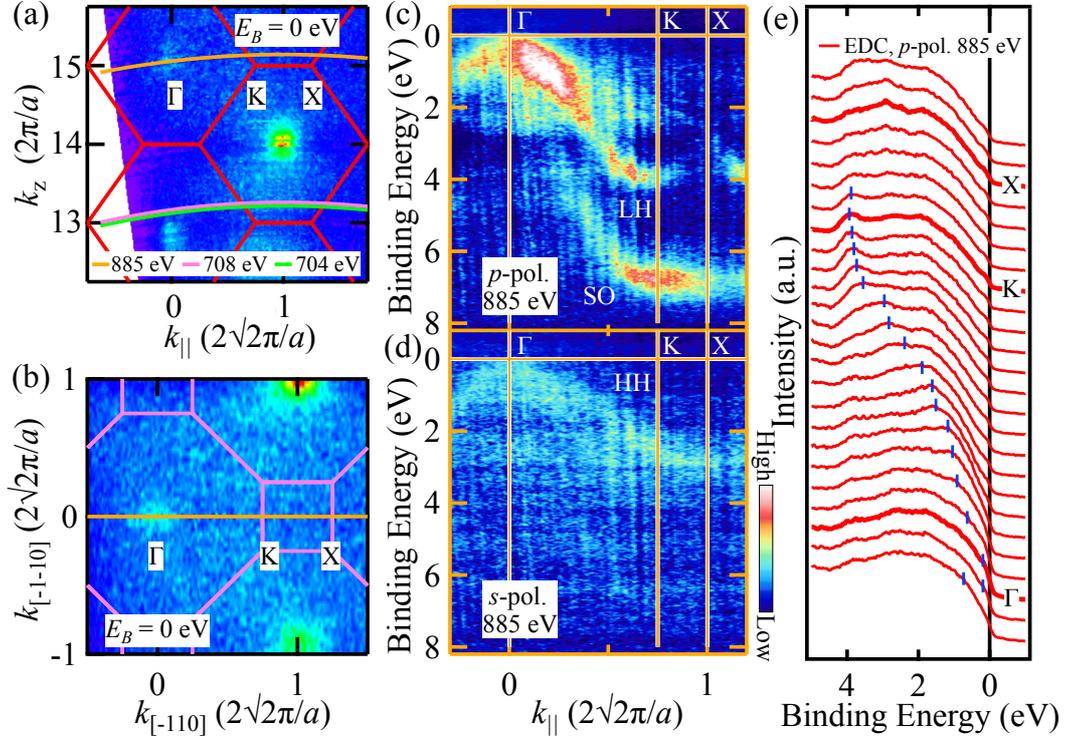

FIG. 3: Band dispersion around the Γ point of a $(Ga_{0.95},Fe_{0.05})Sb$ thin film: (a) FSM in the $k_z$-$k_\parallel$ plane where the yellow curve represents the cut at $h\nu = 885$ eV, and the red solid lines represent the BZ; (b) FSM in the $k_{[-110]}$ - $k_{[-1-10]}$ plane taken at $h\nu = 885$ eV where purple solid lines represent the BZ; (c), (d) ARPES images along the Γ-K-X line taken with $p$ and $s$ polarizations, respectively. The spectra are measured at $h\nu = 885$ eV, and the LH, HH, and SO denote the light-hole, heavy-hole and split-off bands, respectively; (e) EDCs along the Γ-K-X line corresponding to the ARPES spectrum shown in panel (c), and the blue vertical bars are a guide to the eyes tracing the band dispersion of the LH band.

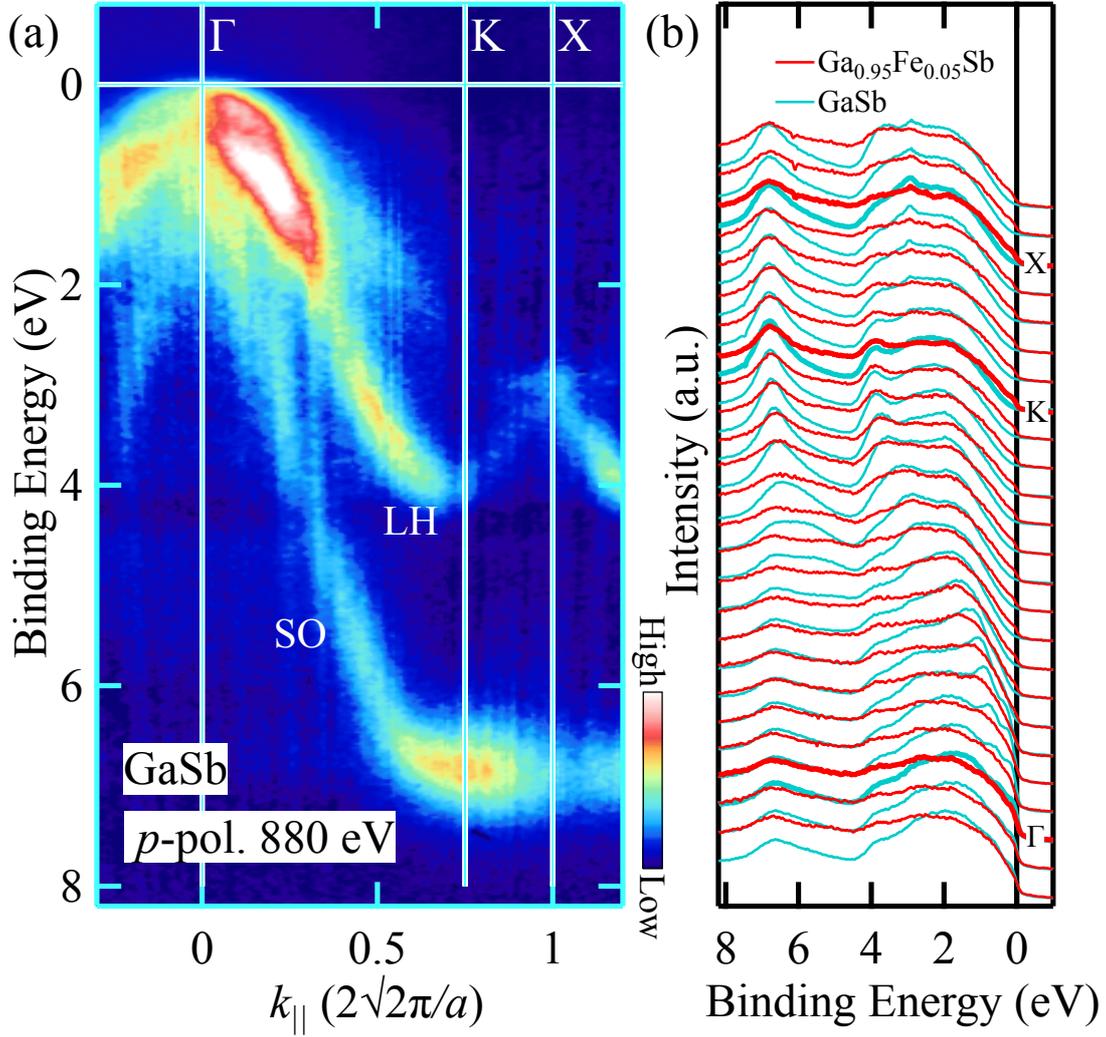

FIG. 4. Band dispersion near the Γ point of GaSb; (a) ARPES image of GaSb along the Γ-K-X line measured with *p* polarization at *hv* = 880 eV, (b) EDCs of the (Ga$_{0.95}$,Fe$_{0.05}$)Sb (red curves, from Fig. 3(c)) and GaSb (blue curves, from Fig. 4(a)) thin films.

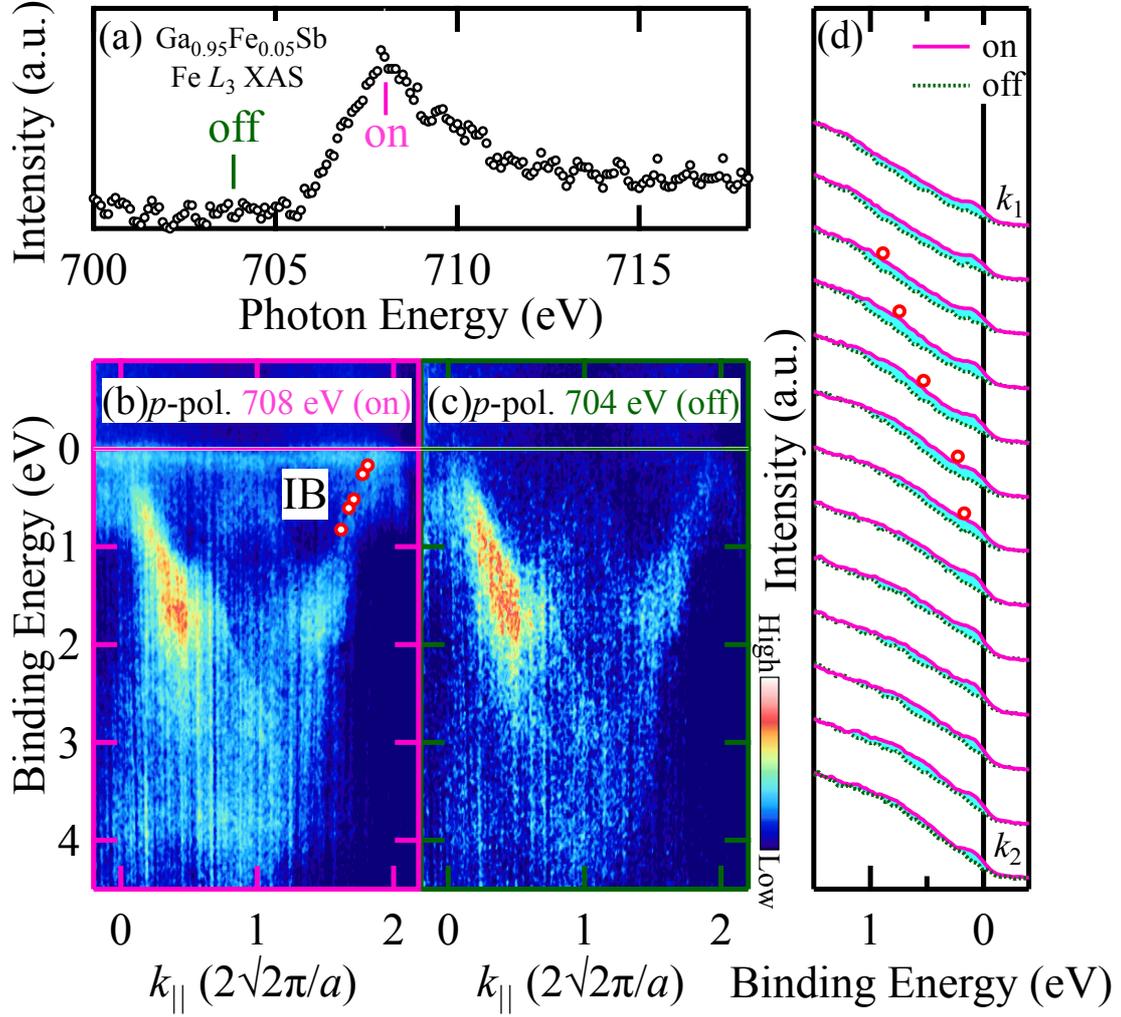

FIG. 5. Resonant ARPES spectra of $(Ga_{0.95},Fe_{0.05})Sb$: (a) Fe $L_3$ XAS spectrum for $h\nu$ = 708 eV and 704 eV are for the on-resonance and off-resonance, respectively; (b), (c) on- and off-resonant ARPES images measured with $p$ polarization, respectively; (d) EDCs of the on-resonant ARPES (pink solid curves, from Fig. 5(b)) and off-resonant ARPES (green dashed curves, from Fig. 5(c)) between $k_{||}$ =1.5 ($k_1$) and $k_{||}$ = 2.1 ($k_2$). Shaded (blue) area denotes the Fe-$3d$ component of ARPES spectra. Red dots on pink solid curves in Figs. 5(b) and 5(d) represent the points where the on-resonant intensity is remarkably stronger than the off-resonant intensity.

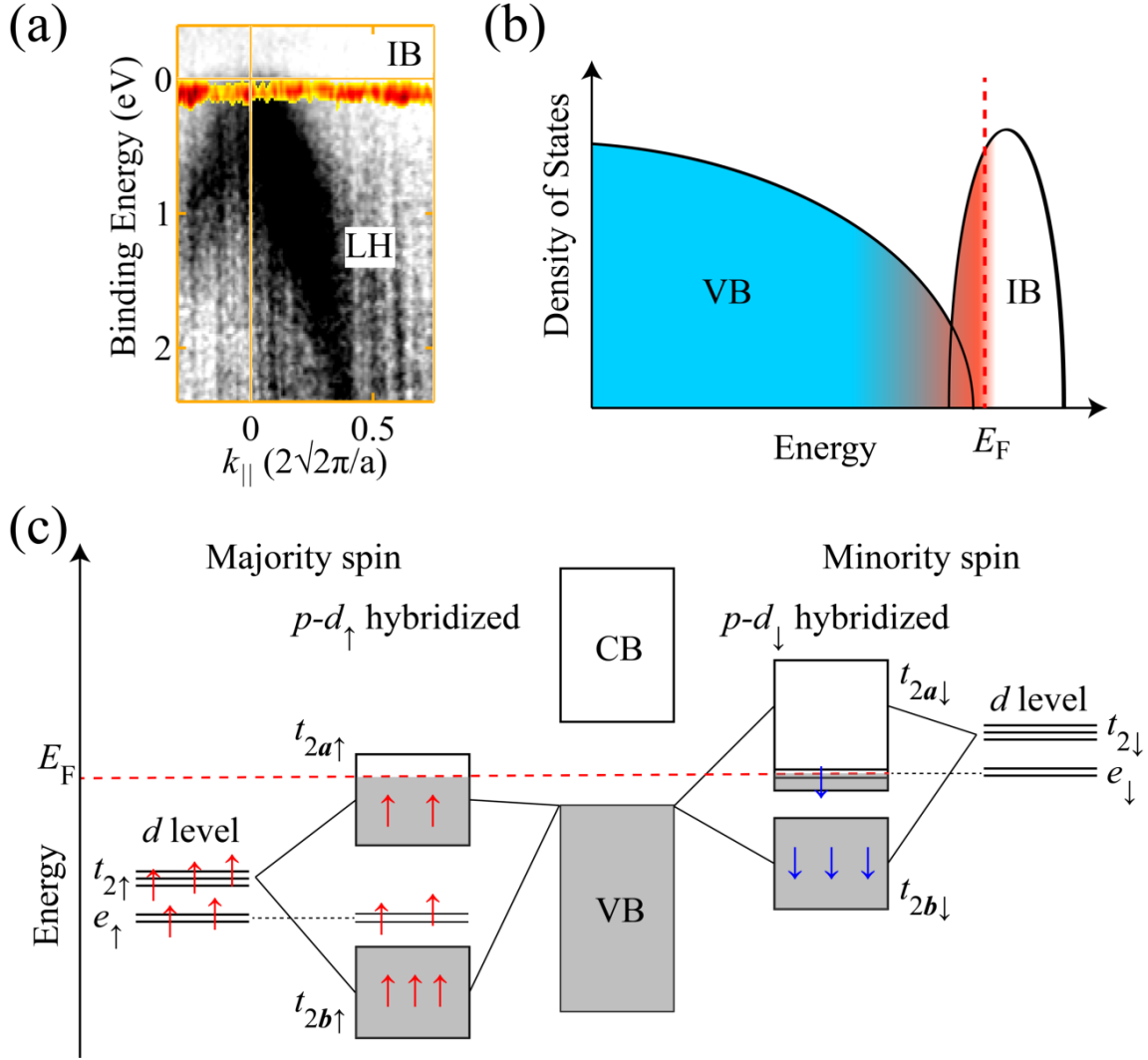

FIG. 6: Electronic structure of $(Ga_{0.95},Fe_{0.05})Sb$: (a) ARPES image from Fig.3(c) with added Fe-3$d$ IB in the vicinity of $E_F$ of Fig. 5(b). (b) Schematic diagram for the density of states (DOS) of $(Ga_{0.95},Fe_{0.05})Sb$. (c) Schematic energy diagram of $(Ga_{0.95},Fe_{0.05})Sb$ where the $p$-$d$ hybridization splits $t_2$ states into antibonding ($t_{2a}$) and bonding ($t_{2b}$) states. CB means the conduction band.